\documentclass[doublecol]{epl2} 

\newcommand{\eqref}[1]{(\ref{#1})}

\renewcommand{\revision}{}

\title{Reconstruction of a scalar voltage-based neural field network from observed time series}
\shorttitle{Reconstruction of a neural network} 

\author{A. Pikovsky\inst{1,2} }
\shortauthor{A. Pikovsky}

\institute{                    
  \inst{1} Institute for Physics and Astronomy, 
University of Potsdam, Karl-Liebknecht-Str. 24/25, 14476 Potsdam-Golm, Germany\\
  \inst{2} Research Institute for Supercomputing, Nizhni Novgorod State University,
Gagarin Av. 23, 606950, Nizhni Novgorod, Russia
}

\abstract{We present a general method for reconstruction of a network of nonlinearly coupled neural fields from the observations. 
A prominent example of such a system is a dynamical random neural network model studied by  
Sompolinsky et. al [Phys. Rev. Lett., v. 61, 259 (1988)]. We develop a technique for  inferring the properties 
of the system from the observations of the chaotic voltages. Only the structure of the model is assumed to be known, while
the nonlinear gain functions of the interactions, the matrix  of the coupling constants, and the time constants of the local
dynamics are reconstructed from the time series.
}

\begin{document}

\maketitle


\section{Introduction}
Reconstruction of networks based on the observation of their dynamics
is a challenging problem relevant for many 
interdisciplinary applications  in physics, climate system analysis, 
biochemical and biological dynamics, genetic regulation, 
epidemiology~\cite{Li_etal-11,Siguhara_atel-12,Oates_etal-14,Tomovski-Kocarev-15,Banos_etal-15,Hirata_etal-16,Tirabassi_etal-17}
and even in social sciences~\cite{Volpe_etal-16}.
Particularly broad are applications in neurosciences, aimed
at an understanding of brain connectivity and 
functionality~\cite{Dickten-Lehnertz-14,Maksimov_etal-14,Pastrana-13,Sporns-13,Boly_et_al-12}. Here
one tries to reconstruct the interactions between the nodes exploring
multivariate neurophysiological 
measurements~\cite{Lehnertz-11,Chicharro-Andrzejak-Ledberg-11,Yu-Parliz-11}.

Generally, methods of reconstruction can be divided in two classes. In the first approach,
one explores statistical interdependencies of observed stochastic processes, and calculates
cross-correlations and mutual (Granger) information measures~\cite{Lusch_etal-16,Schelter-Timmer-Eichler-09,Andrzejak-Kreuz-11,Yang_etal-17}. In another class of methods, one assumes a complex 
dynamical system behind the observations, and tries to reconstruct the network on the basis of the deterministic
dynamics~\cite{Shandilya-Timme-11,Kralemann-Pikovsky-Rosenblum-11,%
Kralemann-Pikovsky-Rosenblum-14,Sysoev_etal-14,Levnajic-Pikovsky-14,Sysoev_etal-16}.

Here we propose a dynamics-based 
method for reconstruction of a neuronal network from the observed voltages
of the nodes. We assume that the dynamics is governed by a generic system of coupled neural fields,
where all the elements - the gain functions, the time constants, and the coupling constants of the interaction are unknown.
In the course of the reconstructions, based on the multivariate time series of voltages, 
these parameters and functions are inferred.    

\section{Neural Network Model and its Dynamics}
\label{sec:nnm}
In this paper we focus on the reconstruction of the network structure that governs neural fields
in the voltage formulation, one of the basic models
in computational neuroscience (see 
Refs.~\cite{Hoppensteadt-Izhikevich-97,Bressloff-12,Ermentrout-Terman-10}). 
Each of the $n$ nodes is characterized by its time-depending
voltage $x_j(t)$, the evolution of which
is governed by the inputs from other nodes according to a system of ordinary differential equations
\begin{equation}
\frac{dx_j}{dt}+\gamma_j x_j=\sum_{k=1}^n C_{jk} F_k(x_k), \quad j=1,\ldots,n\;.
\label{eq:netw}
\end{equation}
Here $\gamma_j$ is the time constant of relaxation of the field at node $j$, and $F_k$ are
the gain functions at the nodes (typically $F(x)\sim\tanh(x)$ or have some similar form). The network is determined by the $n\times n$ coupling
matrix $C_{jk}$. As has been shown in Ref.~\cite{Sompolinsky88}, at strong enough coupling
such a network demonstrates chaos, and this is a state 
which allows  the reconstruction
of
the network matrix $C_{jk}$, the time constants $\gamma_j$, and the functions $F_k$ 
from the set of observations $x_j(t)$, as described below (see Conclusion for a possible 
extension to non-chaotic states).

For our approach the most important  property of the model \eqref{eq:netw}  is its scalar character:
the dynamics at each node is one-dimensional, so it is fully characterized  by a scalar variable $x_k$,
and the nonlinear function $F$ is a function of one variable. This should be contrasted to more general networks
where the dynamics at each node is high-dimensional. 

\begin{figure}[!hbt]
\centering
\includegraphics[width=0.9\columnwidth]{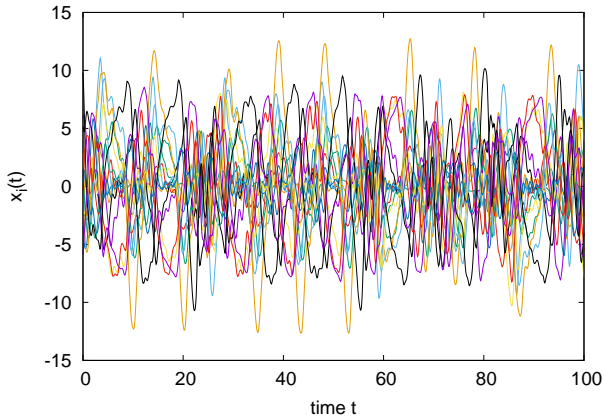}
\caption{(color online) Example of chaotic neural fields $x_i(t)$ in model \eqref{eq:netw}.}
\label{fig:fields}
\end{figure}

First,
we illustrate a chaotic state in system \eqref{eq:netw}. 
To be as close as possible to the theoretical approach of Ref.~\cite{Sompolinsky88}, we take $F(x)=\tanh(x)$ and choose
$C_{ij}$ to be independent random variables with a Gaussian distribution 
with zero mean and standard deviation
$2$ \revision{(the standard deviation is an essential parameter in theory \cite{Sompolinsky88},
chaos in a network is typically observed for large enough values of this parameter)}. The time constants $\gamma_j$ are chosen from a uniform distribution in the range $0.8<\gamma<1.2$. An example
of a chaotic regime for an ensemble of $N=16$ elements is presented in Fig.~\ref{fig:fields}. The calculated
largest Lyapunov exponent is $0.22$. This chaotic state is used below in all calculations for the
illustration of the reconstruction method.

\section{Reconstruction of the Network Parameters}
\label{sec:rcm}
\subsection{Method of the reconstruction: known time constants}
Here we present the method of the reconstruction, assuming that the time constants $\gamma_j$ are known.
We will extend to the case of unknown constants $\gamma_j$ in the next subsection.

Suppose one observes the time series of all variables $x_j(t)$ governed by Eq.~(\ref{eq:netw}). 
The problem is to reconstruct the coupling matrix $C$ and the functions $F_k$
from these observations. First, we calculate the time derivatives of the observed signals to obtain
the time series $(\dot x_j, x_j)$. Let us invert the equations \eqref{eq:netw}:
\begin{equation}
F_j(x_j)=\sum_i W_{ji}(\dot x_i+\gamma_i x_i)\;.
\label{eq:inv}
\end{equation}
Here $W=C^{-1}$ is the matrix inverse to the coupling matrix $C$. This matrix generally exists, as the random matrix
$C$ is typically non-singular. 

The main idea of the reconstruction follows from the functional relation of the scalar variable $x_j$ and the
r.h.s. of Eq.~\eqref{eq:inv}. For the sake of simplicity of presentation we denote $y_i(t)=\dot x_i+\gamma_i x_i$.
Suppose that in the chaotic time series $x_j(t)$, we find two time instants, $t_1$ and $t_2$,
such that $x_j(t_1)\approx x_j(t_2)$. Then, from Eq.~\eqref{eq:inv} it follows that $\sum_i W_{ji}y_i(t_1)\approx \sum_i W_{ji}y_i(t_2)$. 
\revision{Here, in fact, only continuity of the function $F_j$ is used; no other assumptions are needed.} This
relation can be rewritten as 
\begin{equation}
\sum_i W_{ji} z_i\approx 0,\qquad z_i=y_i(t_1)-y_i(t_2)\;.
\end{equation}
We expect that for a chaotic time series this relation is non-trivial, i.e. the vector $z_i$ does not vanish. (For a periodic time series
one can obviously find such points that $y_i(t_2)\approx y_i(t_1)$, if $t_2-t_1$ is a multiple of the period. \revision{As discussed 
in Conclusions below, only periodic regimes with a complex waveform may provide enough non-trivial recurrent 
points to ensure reconstruction.)} 

For a sufficiently long scalar time series $x_j(t)$, we in fact can find many such time instants where $x_j(t_1)\approx x_j(t_2)$, and
correspondingly we have a large set of $M$ vectors $z_i^{(m)}$, $m=1,\ldots,M$. \revision{The size of the set $M$ is the essential 
parameter of the method, as dicussed below it is close to the length of the observed time series.} Using a vector notation 
$\{z_i^{(m)}\}=\vec{z}^{(m)}$, we thus
obtain a set of equations
\begin{equation}
\vec{w}_j \cdot \vec{z}^{(m)}=0\;,
\label{eq:svd}
\end{equation}
where we denoted the $j$-s row of the matrix $W$ as a vector: $[\vec{w}_j]_i=W_{ji}$. 

The problem of finding the vector $\vec{w}_j$ from the set of linear equations \eqref{eq:svd}
is the standard problem of finding a null space of the $N\times M$ matrix composed of $M$ vectors 
$\vec{z}^{(m)}$. This problem can be straightforwardly solved
via Singular Value Decomposition (SVD)~\cite{Trefethen-Bau-97}.
Once the zero singular value is found, the corresponding entry in the 
obtained unitary matrix gives the vector $\vec{w}_j$. Performing this for different rows $j$ allows us to
obtain the matrix $W$. We note that the matrix is obtained unambiguously up to a normalization, because 
the functions $F_j$ in \eqref{eq:inv} are unknown. If one assumes these functions to be normalized, then the
inverse coupling matrix $W$ is defined in a unique way. Together with the time series $y_i(t)$, this matrix defines
according to Eq.~\eqref{eq:inv} the gain functions $F_j$. Finally, the coupling matrix $C$ is obtained by inversion of $W$
(\revision{we remind that the time constants $\gamma_j$ are assumed to be known in this variant of the method}).
Practically, one obtains not exact null spaces, but  spaces corresponding to the minimal singular values 
$S_j^{(min)}$,
which are small but do not vanish exactly.

In the procedure above we have to find close returns of the time series $x_j(t)$. Because this time series is a scalar one,
the following simple technique could be used. One just performs a sorting of the available array of values $x_j$. Then
the neighboring values of $x_j$ in the sorted array \revision{(although they are of course typically not neighbors in the original time series
if the sampling rate is not too large; for a large sampling rate
one needs to exclude neighbors within, say, a typical correlation time in the original time series)}
provide the closest returns. If the range of values of $x$ is $\Delta$, then for $M$ points in the chaotic time series
one can estimate $|x_j-\hat{x}_j|\sim \Delta/ M$, where $\hat{x}_j$ is a close neighbor of $x_j$ after sorting. One can see 
that the error vanishes for a very long time series $M\to\infty$. Furthermore, taking only nearest neighbors in the 
sorted time series, one avoids redundancy in the matrix of vectors $\vec{z}^{(m)}$, as each value of $x_j$ participates
only twice in the formation of the set of $\vec{z}^{(m)}$.

\begin{figure}[!hbt]
\centering
\includegraphics[width=0.9\columnwidth]{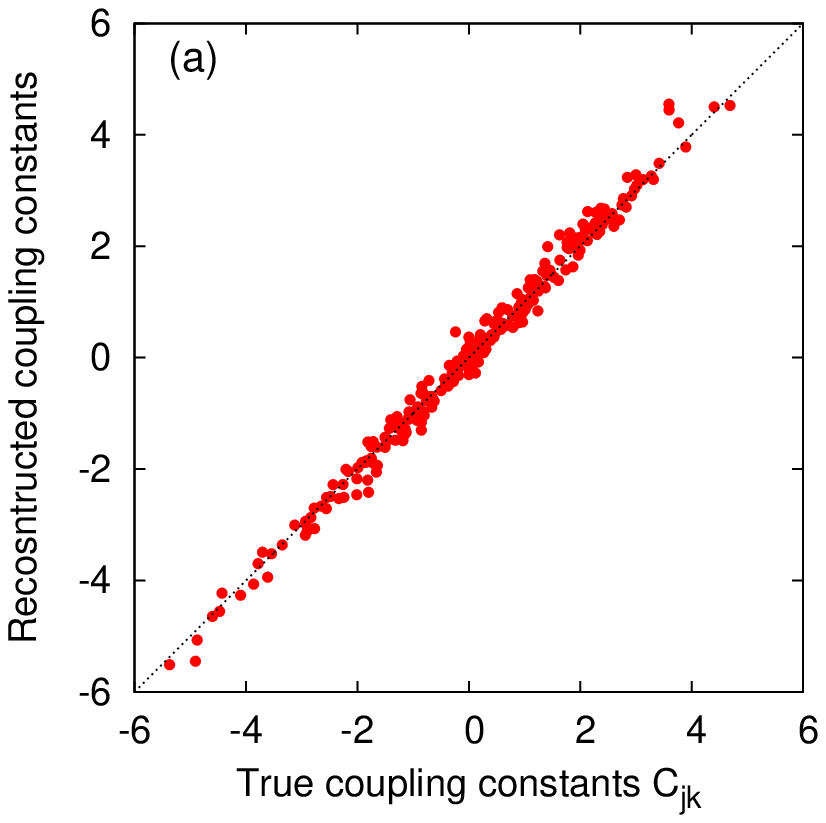}
\includegraphics[width=0.9\columnwidth]{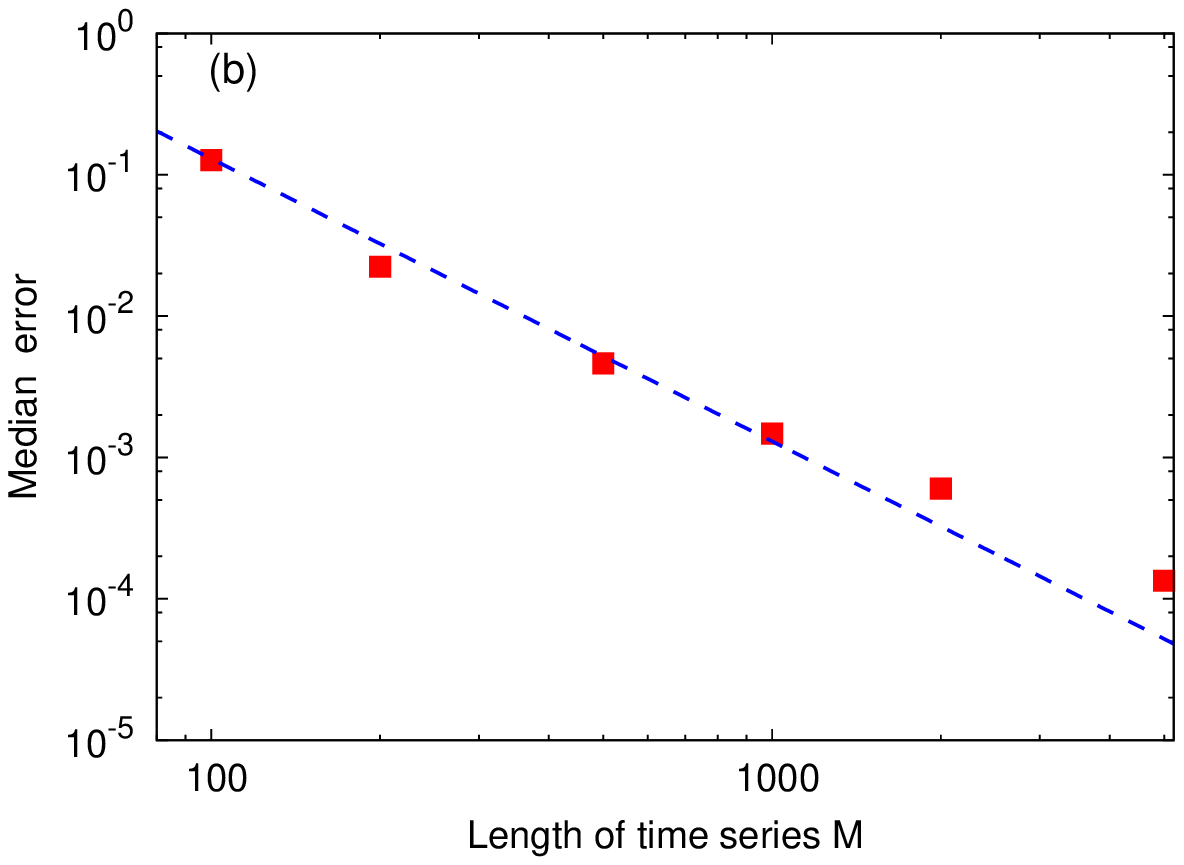}
\caption{Reconstruction of the coupling constants. Time derivatives are calculated via the Savitzky-Golay filter with parameters
$[6,6]$, for the time step $dt=0.01$. Points for the analysis were taken with step $\Delta t=2$. Panel (a): reconstructed vs 
true coupling constants  for $M=100$. Dotted line is the diagonal. Panel (b): dependence of the median error on the length of the time series $M$ (square markers).
The dashed line has slope of $-2$.}
\label{fig:rec1}
\end{figure}

\begin{figure}[!hbt]
\centering
\includegraphics[width=0.9\columnwidth]{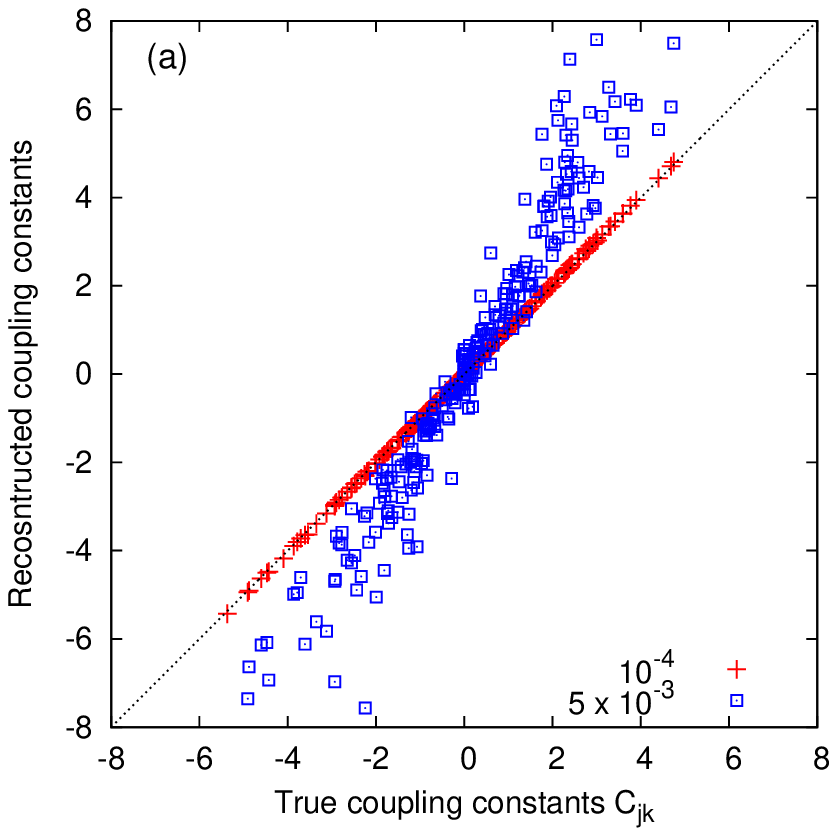}
\includegraphics[width=0.9\columnwidth]{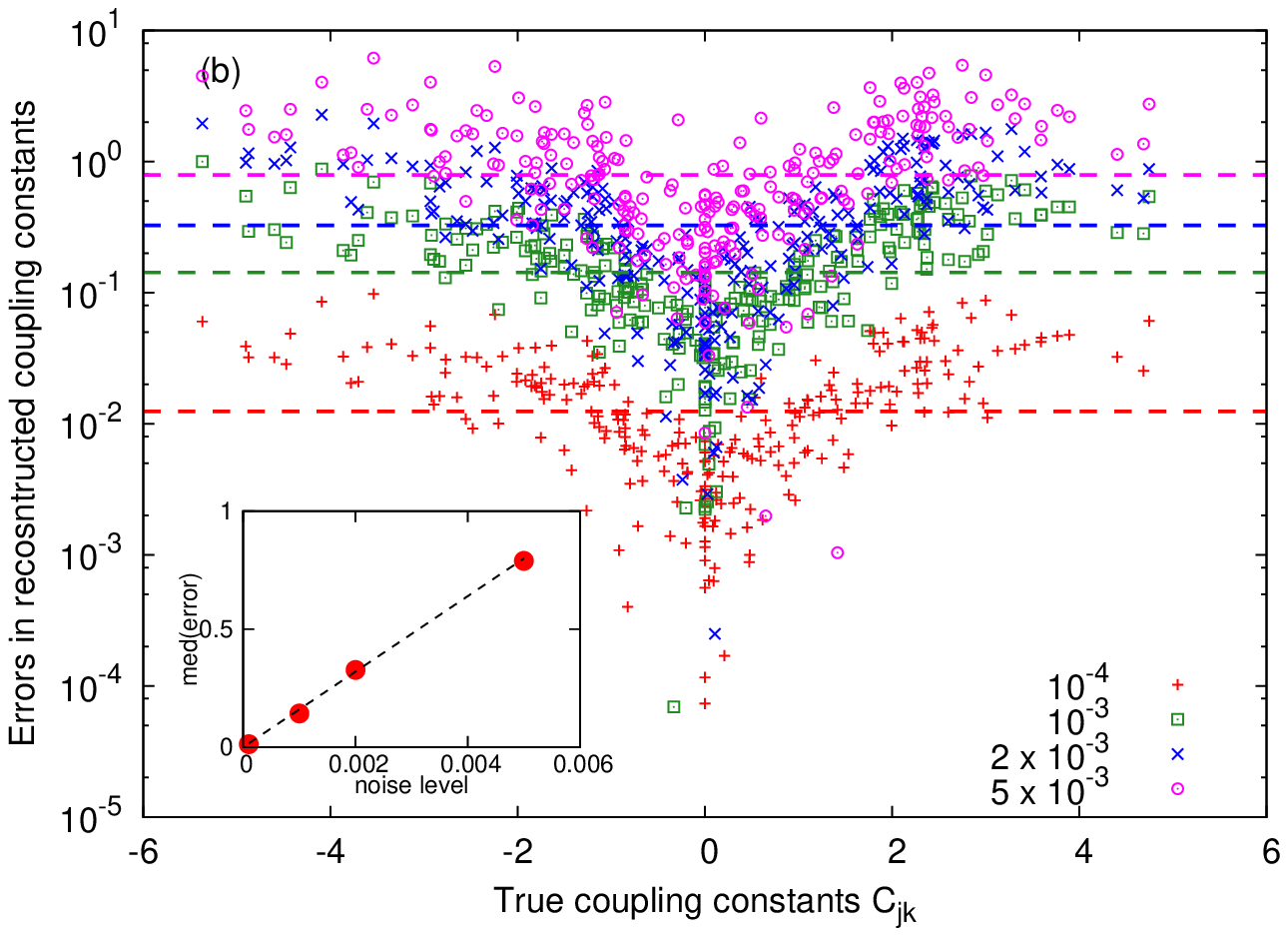}
\caption{\revision{Reconstruction of coupling constants in presence of noise. Time derivatives are calculated 
via the Savtzky-Golay filter with parameters
$[6,6]$, for the time step $dt=0.01$. Points for the analysis were taken with step $\Delta t=2$, the total number of points was
$M=1000$. Four levels of noise have been explored. Panel (a) shows the reconstructed vs true 
coupling constants for $\sigma=10^{-4}$ (pluses, nearly perfect reconstruction for the minimal noise level) 
and $\sigma=5\cdot10^{-3}$ (squares, rather large level of errors). Dotted line is the diagonal.
Panel (b) shows all the errors in the reconstructed  coupling constants vs the true ones, for  four indicated levels of noise.
Dashed lines show the medians of these errors. The inset in panel (b) depicts linear dependence of the median of the errors
vs noise level (dashed straight line has slope $160$).}
}
\label{fig:rec2}
\end{figure}

We illustrate the method in Fig.~\ref{fig:rec1}. Here we used a multivariate time series $x_j(t)$ depicted in Fig.~\ref{fig:fields},
from which the time
derivatives have been calculated by virtue of the Savitzky-Golay filter. The panel (a) shows the reconstructed coupling constans
of the matrix $C$ vs. the original ones, for a rather small length of the time series $M=100$. The panel (b) shows the dependence
of the median error on the length of the time series $M$ \revision{(we use median because of a broad distribution of errors,
cf. Fig.~\ref{fig:rec2}(b) below)}. This dependence fits quite well the scaling $Err \sim M^{-2}$.

To check the robustness of the method to different levels of noise in the data, we performed the same analysis 
as shown in Fig.~\ref{fig:rec1} for the time series with
added observational Gaussian noise with standard deviation $\sigma$. Partially this noise is filtered out due 
to application of the Savitzky-Golay 
filters needed also to calculate the time derivatives. Figure~\ref{fig:rec2} illustrates the 
quality of the reconstruction for four different values of $\sigma$. \revision{One can see that the 
errors are proportional to the noise level. 
The reconstruction is, nevertheless, quite robust as even for a large noise level in Fig.~\ref{fig:rec2}(a), 
the reconstructed coupling
constants
though inaccurate in absolute values, are nevertheless roughly linearly proportional to the true values (cf. squares in
Fig.~\ref{fig:rec2}(a)) and this allows
distinguishing strong and week links in the network.}

\subsection{Unknown time constants}

Above we have assumed that the time constants $\gamma_j$ at the nodes are known. This allowed us
to calculate the values $y_i(t)$ explicitely. In the case of unknown $\gamma_j$ this is not possible, and to apply
the method above we have to scan over different test values of $\gamma_j$. An indicator for the correct choice 
of the time constants is the quality of the found null space of the system of equation~\eqref{eq:svd}. Practically, we used the
maximum over index $j$ singular value
\begin{equation}
S(\vec{\gamma})=\max_j S_j^{(min)}
\end{equation}
as the cost function, trying to minimize it.

\begin{figure}[!hbt]
\centering
\includegraphics[width=0.8\columnwidth]{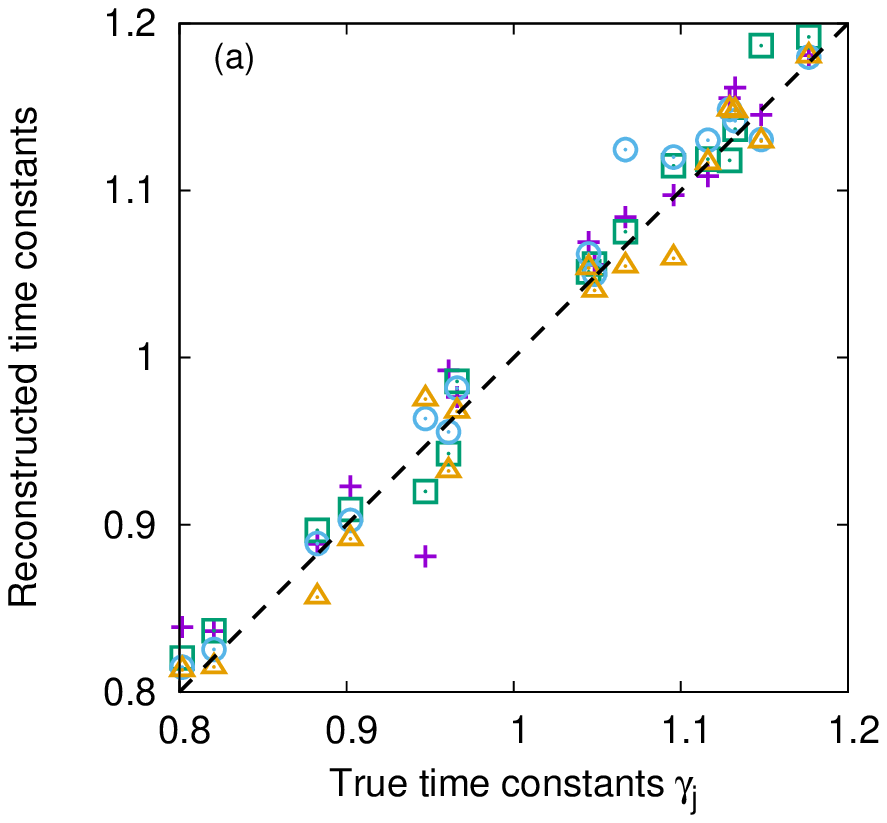}\\
\includegraphics[width=0.9\columnwidth]{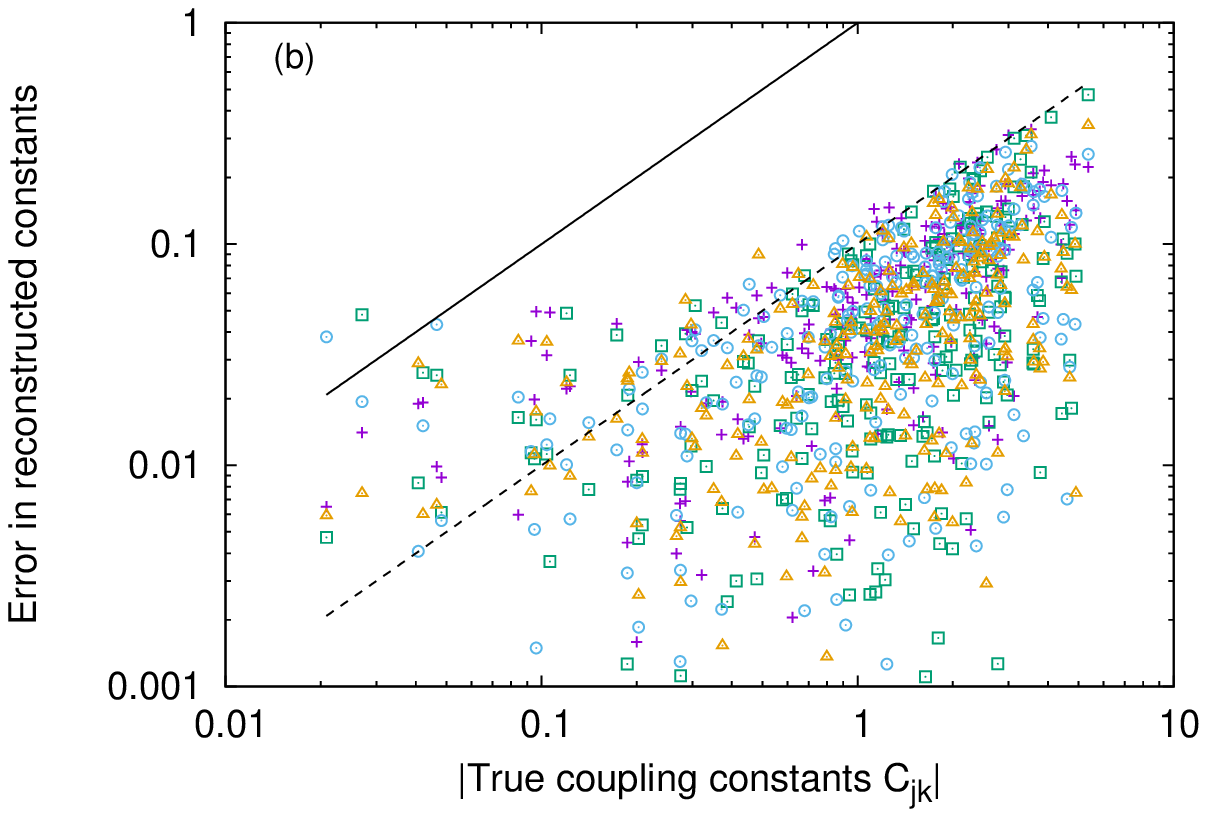}
\caption{Reconstructed time constants (a) and the error $Er=|C_{rec,jk}-C_{jk}|$ in the
reconstructed coupling constants 
vs. the true constants themselves (b). 
Four symbols show four independent runs 
of the simulated annealing routine. Number of points used $M=1000$.
In panel (b) two lines show levels $Er=C$ and $Er=0.1 C$ (dashed line). As 
most of the values lie below the dashed line, the maximal 
error of the method can be estimated as 10\%, except for coupling constants 
that are very small - for them the relative error is of order 1.}
\label{fig:rec3}
\end{figure}

Unfortunately, in the definition of
vectors $y_i$ and $\vec{z}^{(m)}$ all the time constants enter, so one has to scan in the full $N$-dimensional space to 
find an absolute minimum of $S(\vec{\gamma})$.
Thus, \revision{because a brute force approach is hardly possible,}  a more sophisticated search is needed. We have found that a simulated annealing approach 
(see \cite{Press-Flannery-Teukolsky-Vetterling-89}, Chapter 10), provides a proper estimate of the time constants. 
In our realization we attributed $\log S(\vec{\gamma})$ to the ``energy'' and decreased the 
``temperature'' of the simulated annealing
from $0.1$ to $5\cdot10^{-3}$, multiplying it at each step by $0.999999$. The variations of the
vector $\vec{\gamma}$ have been performed by
adding at each step a random Gaussian vector with amplitude $0.05$. To estimate 
reliability of the procedure, we performed four 
runs, checking if the same values of the time constants $\vec{\gamma}$ and the 
same values of the coupling matrix $C$ are
reached in these independent implementations of the annealing. The results for a time series
of length $M=1000$ are shown in 
Fig.~\ref{fig:rec3}. One can see that the reconstruction is not perfect,  however, the likelihood that the relative error 
for the coupling constants that are larger than $0.1$ is less than 10\%.

\section{Conclusions}
In summary, we have developed a method to reconstruct the connections of the network behind
a collection of interacting neural fields, provided the observations of the 
potentials at the nodes are available. The method delivers the connectivity matrix, together with
the parameters (time constants) characterizing the dynamics of the nodes, and the nonlinear gain functions defining 
the interactions. No prior knowledge on any of these parameters
is required, only the general structure of the system is supposed to be known.
We have assumed that
data for all the nodes are available; exploration of the situation with unobserved nodes
 is a subject of ongoing research.
 
 Our method heavily relies on the diversity of the dynamics, and works most powerfully 
 in the case the dynamics is chaotic. Additional tests have shown that even for a periodic
 dynamics, if this is complex enough (i.e. the wave form is nontrivial with several minima and maxima
 over the period) to ensure sufficient diversity, the reconstruction works well. However,
 if the dynamics is periodic with a simple sine-type waveform, the reconstruction fails as 
 the vectors used in the SVD analysis are not independent. 
 
 It is worth comparing our approach with other methods of dynamical
 network reconstruction. The most similar approach is that of Ref.~\cite{Pikovsky-16},
 where a neural network model in the firing rate formulation was considered.
 The difference to the present work is that in the firing rate formulation instead
 of Eqs.~\eqref{eq:netw} one has a system
 \begin{equation}
\frac{dy_j}{dt}+\gamma_j y_j=G_j\left(\sum_{k=1}^n C_{jk} y_k\right), \quad j=1,\ldots,n\;.
\label{eq:frnetw}
\end{equation}
Because of the different order of applying a nonlinear function $G$ and a linear coupling, to treat system
\eqref{eq:frnetw} one has to invert the gain function $G$, instead of inverting the coupling 
matrix $C$. Therefore, the approach of Ref.~\cite{Pikovsky-16} applies to invertible gain 
functions only, while in the present case no restriction on the nonlinear functions in \eqref{eq:netw}
is imposed -- \revision{instead here the coupling matrix has to be non-singular, what appears
to be a typical case for random matrices}.
 In Ref.~\cite{Shandilya-Timme-11} a general setup of high-dimensional
 dynamical systems coupled via a network of nonlinear interactions was considered. It was assumed that 
 all nonlinear functions defining the local dynamics and the coupling are known, so the only unknown
 parameters, the coupling constants, could be reconstructed provided the full high-dimensional time series
 from all sites are available. Similar assumptions have been made in 
 Refs.~\cite{Levnajic-13,Levnajic-Pikovsky-14,Leguia_etal-17}. In the present work, no knowledge 
 on the nonlinear coupling 
 functions is required. The local dynamics is assumed to be linear, 
 thus only one unknown parameter (time constant) has to be determined at each node. 
 In Ref.~\cite{Wang-Lai-Grebogi-16} applications of compressive sensing methods to network 
 reconstruction are reviewed. In these methods one assumes the network to be sparse, while no
 such assumption is needed for our approach.

\acknowledgments
We acknowledge useful discussions with M. Rosenblum and I. Sysoev.  Results from Sects. 2-3 
were supported by the Russian Science Foundation (Contract No. 17 12 01534).

\def\cprime{$'$}

\end{document}